# AFM-compatible near-field scanning microwave microscope with separated excitation and sensing probes

K. Lai, M.B. Ji, N. Leindecker, M.A. Kelly, and Z.X. Shen

Dept. of Applied Physics and Geballe Laboratory for Advanced Materials, Stanford University, Stanford, CA 94305

**Abstract**

We present the design and experimental results of a near-field scanning microwave microscope (NSMM) working at a frequency of 1GHz. Our microscope is unique in that the sensing probe is separated from the excitation electrode to significantly suppress the common-mode signal. Coplanar waveguides were patterned onto a silicon nitride cantilever interchangeable with atomic force microscope (AFM) tips, which are robust for high speed scanning. In the contact mode that we are currently using, the numerical analysis shows that contrast comes from both the variation in local dielectric properties and the sample topography. Our microscope demonstrates the ability to achieve high resolution microwave images on buried structures, as well as nano-particles, nano-wires, and biological samples.

## I. Introduction

Much of our understanding of the physical properties of a material comes from its interaction with electromagnetic waves. In the far-field radiation region, the measured electrodynamic property is averaged over a length scale in the order of the wavelength ($\lambda$), which limits the spatial resolving power to ~ $\lambda/2$, known as the Abbe barrier. In the near-field region with $D$ (dimension of the source) and $r$ (source-sample distance) much smaller than $\lambda$, however, the time-oscillating electric and magnetic fields strongly depend on the surrounding materials and contain information of the local sample properties. Combined with serial pixel-to-pixel data acquisition techniques, the spatial resolution of scanning near-field microscopy [1] is then set by the spatial extent of these evanescent fields, rather than the free-space wavelength.

After the original proposal by Synge [2], near-field microscopy was first demonstrated at microwave frequencies [3] and then extended to the far infrared, infrared, and visible regimes [1]. The past decade has seen renewed effort in the microwave regime to develop a near-field scanning microwave microscope (NSMM) as a scientifically useful instrument [1,4]. An NSMM should in principle allow one to directly probe the local radiofrequency (RF) electrodynamic properties, e.g., the surface impedance ($Z_S$), the complex dielectric permittivity ($\varepsilon$), and permeability ($\mu$) [5]. Such a unique feature can be utilized to study both fundamental physical processes, such as classical and quantum phase transitions, and applied science, such as the electrical properties of biological tissues and biopolymers at the microscopic level. On the practical side, state-of-the-art microelectronics and telecommunication devices all operate at GHz frequencies, creating huge practical interest in understanding material properties in the microwave regime. Finally, compared to the atomic force microscope (AFM) and scanning tunneling microscope (STM), the long-range electrostatic-like force involved in the NSMM relieves the stringent requirement of proximal probes, thus enabling high-speed, non-contact, and non-destructive measurements [1,4,5].

The earliest demonstration of the NSMM utilized microwave cavities with a small aperture [3,6]. This geometry suffers strong attenuation because a small aperture is effectively a waveguide working beyond the cutoff condition. The cutoff problem was circumvented by a second design of NSMM based on transmission lines [7,8]. Further improvements incorporated a resonant structure at the probe end of the transmission line [9]. Monitoring the change of resonant frequency and quality factor (Q), several groups have demonstrated different designs of resonant NSMMs with very high sensitivity and spatial resolution [9-13].

Despite the encouraging progress in NSMMs, several limitations exist in current designs [14]. First, for the single transmission line probe, either resonant or non-resonant, the excitation and the sensing unit share the same electrode. The voltage level in the signal line, which is termed here the common-mode signal, is inevitably high, resulting in large shot noise. Second, in several designs with a feedback mechanism [4], the reflected wave is used for both detection and tip-sample distance control, thus no independent signal exists for positioning. Finally, the sensing unit of many resonator-based NSMMs is bulky and the sharp STM-like tips are fragile [12-14]. This leads to slow operation and low bandwidth, compromising the unique nature of long-range force involved in a microwave microscope.

In this paper, we present a new design using two transmission lines, one for excitation and the other for sensing. With a proper design, the detector can be "orthogonal" in that the signal is minimized in the absence of a sample. The ultimate goal is to implement a high throughput NSMM with isolated excitation and sensing electrodes, both integrated onto a miniature and flexible tip, while still maintaining high sensitivity and spatial resolution.

## II. System design

**A. Sensor design**

The key feature of our NSMM lies in the design of an orthogonal probe that fulfills the above requirements. We present a detailed design of our sensor built onto an AFM cantilever by standard micro-fabrication processes, taking advantage of a well developed platform [15,16] that provides the scanning and distance control functions.

Fig. 1(a) shows the picture of the overall tip design, which replaces a standard AFM tip assembly, and includes the two transmission lines. The scanning platform we are using is a customized AFM Scanner (Pacific Nanotechnology, Inc., Santa Clara, CA). Two 1μm-thick aluminum coplanar waveguides (CPWs) with 50Ω characteristic impedance are patterned on the bulk part of a silicon chip. One end of each CPW is connected to a bonding pad, and the other end extends all the way to the cantilever beam. The silicon nitride ($Si_3N_4$) cantilever is 500μm long and 2μm thick, and is formed by KOH back-etch of the Si substrate. We chose $Si_3N_4$ as the cantilever material because it is less lossy at GHz frequencies than Si, and more robust than other insulators like $SiO_2$.

Fig. 1(b) shows the mask design near the end of the cantilever (right) and the schematic of its cross-section. The first metallization is simply a short Al strip on top of the $Si_3N_4$ insulator. The end of the cantilever beam is then covered by depositing a layer of 3μm $SiO_2$, with a 3μm-diameter hole in the center. The top Al layer forms all three (excitation, sensing, and ground) electrodes on the entire chip. Note that while the sensor line for the second Al layer stops at the edge of the $SiO_2$ coating, the sensing electrode extends all the way to the center hole through the buried Al strip. Such a two-layer implementation proves to be much better than a single-layer structure [14] in common-mode isolation and reduces the noise level significantly. Even with this shielding scheme, however, the direct coupling between the excitation and sensing electrodes is still severe and a common-mode cancellation circuit is needed during the microscope operation.

Finally, using a focused-ion beam (FIB) system, we deposited Pt into the center hole for the sensing electrode, as shown in Fig. 1(c). A tip radius of 100nm is easily achieved in the FIB process, which, in principle, should allow us to reach very high spatial resolution. The drawback is that the FIB process is slow and costly, and often compromises the electrical characteristics of the tips. The Pt-tip is also quite fragile during the operation and degrades with use.

**B. Circuit design**

In its essence, our sensing unit measures the transmission from the excitation electrode to the sensor. Signals to and from the probe are routed to 50Ω matching networks adjacent to the probe. These are optimized at the operating frequency of 1GHz, but in principle could be broadband transformers. We emphasize that, unlike many existing NSMMs that primarily utilize the frequency shift and change of the Q-factor for the measurement [1,4], the electronics here measures the in-phase and out-of-phase components of the sensed signal to determine the complex impedance of the sample. .

The microwave circuit diagram is shown in Fig. 2. As mentioned above, due to the incomplete shielding between the excitation and the sensing electrodes, a large common-mode component exists in the sensor signal even in the absence of a sample. To avoid saturating the RF amplifier, the common-mode signal is nulled by injecting a properly phase-shifted and attenuated signal from the generator. In the closed-loop condition, feedback from the output channels can adjust the attenuation in the cancellation unit to correct small drifts of the null signal during scanning. After the cancellation, the detected signal is amplified and demodulated in a quadrature mixer. The in-phase and out-of-phase signals are then amplified and recorded during the AFM scan. After proper calibration in both phase and amplitude, these two output channels manifest the real and imaginary parts of the complex dielectric constant of the sample.

One advantage of having two electrodes in the sensing unit is that other signals like the reflection from either electrode can be utilized for the tip-sample distance control. As shown in Fig. 2, the reflected signal from the excitation electrode is collected and detected by the height servo circuit. The output of this circuit serves as the feedback signal that regulates the z-piezo in the AFM. Due to the large size (~ 20μm in diameter) of the excitation ring, one measures the averaged reflection over this larger area instead of following every detail of the sample topography. Provided that the electrical property does not change dramatically in the large scale, such servo should be able to keep a constant tip height above the sample. Most data in this paper were taken in contact mode with no feedback control, and we are making effort to implement a reliable height servo for non-contact mode.

## III. Numerical analysis

An attractive feature of NSMMs is the ability to image material properties in a quantitative manner [5]. This is largely due to the relatively simple physics when matter interacts with electromagnetic waves in the microwave regime. In our case, since all relevant dimensions are much smaller than the free-space wavelength (30cm) at the working frequency, the wave nature in the near-zone can be ignored and the electrostatic approximation applies.

The finite element analysis (FEA) was performed by using the linear electrostatic model in a software package FEMLAB [17]. As can be seen in Fig. 1, the sensing unit at the end of the cantilever is approximately axisymmetric in shape. As a result, a 2D axisymmetric model was used in most cases to avoid the slow calculation speed and large errors of a large 3D mesh. We believe that the 2D approximation captures the essential physics of the near-field interaction between the sensing unit and the sample.

Fig. 3 shows the FEMLAB results of the equipotential contours close to the sensing unit for the FIB tip. The sharp Pt-tip is in contact with a sample having a relative dielectric constant ($\varepsilon_r$) of 4. The dimensions in this model are taken from the SEM image (Fig. 1(c)) of the micro-fabricated cantilever. For simulation purposes, the Pt-tip is assumed to be 1µm-tall with 50nm radius in the apex. The boundary condition is chosen so that the regions far away from the system (40µm) are grounded. The excitation voltage is set to 1V and the shield grounded.

Fig. 4(a) shows the change of the tip voltage in the presence of a sample, $\Delta V_{tip} = V_{tip}(\varepsilon_r) - V_0$, where $V_0$ is the tip voltage for $\varepsilon_r = 1$ (air). When the tip scans over a flat sample with regions having different dielectric properties, as shown schematically in the inset of Fig. 4(a), this electrical response provides contrast in the output of the microwave microscope. Practically, the electrical signal is generally much smaller because the contrast only comes from local dielectric variation within a thin layer of the sample.

An important feature of the current design is that the tip signal is sensitive to the sample topography, as well as its electrical properties. A quantitative understanding of the topographical response is shown as the approaching curve in Fig. 4(b) when the tip approaches the sample with $\varepsilon_r = 4$. Here the tip signal, $\Delta V_{tip} = V_{tip}(d) - V_0$, is plotted as a function of the spacing $d$ between the excitation electrode and the sample. For our tip geometry, much of the signal appearing on the tip electrode comes from the capacitive coupling between the tip and excitation electrodes, rather than the tip-sample coupling. A sample underneath the sensor acts as a partial shield to the tip-excitation capacitance, resulting in a negative approaching curve [18]. When the tip scans over the rough sample surface in a contact mode, as shown in the inset of Fig. 4(b), the entire sensing unit follows the topography and the signal moves up and down in the approaching curve, resulting in contrast in the microscope output.

## IV. Experimental results and discussions

### A. Electrical contrast

Due to the long-range electromagnetic interaction between microwave and matter, electrical contrast can be obtained even for flat samples with dielectric variations. This is particularly useful for imaging buried structures in the microelectronic integrated circuits, as well as studying electronic phase transition in fundamental physics.

Fig. 5 demonstrates the sub-surface imaging ability of our designed NSMM. To avoid the entanglement of electrical and topographical signals, we prepared a smooth sample (Fig. 5(a)) with sub-surface dielectric contrast as follows. The patterns with 120nm in depth, as seen in Fig. 5(c), are defined in the thick $SiO_2$ layer on top of a silicon wafer. A layer of $Al_2O_3$ is then sputtered onto the device and surface polished. For a consistency check for the smoothness of the surface, the topography is measured by a calibrated Digital Instruments Multimode AFM (Veeco Metrology, Inc., Santa Barbara, CA), as shown in Fig. 5(b). After the polishing, the surface is still fully covered by $Al_2O_3$, as confirmed by X-ray photoemission spectroscopy. Microwave images from the two output channels are shown in Fig. 5(c) and (d). As expected, after properly adjusting the phase ϕ (see Fig. 2), contrast in the conductivity plot vanishes, while the sub-surface dielectric contrast is clearly observed. The contrast signal, about 60mV for this particular sample, can be obtained by taking an arbitrary line cut in Fig. 5(c). A spatial resolution ~ 120nm, consistent with the tip diameter, is also extracted from the rising / falling edge across the sub-surface boundary of two distinct materials. Detailed quantitative results, e.g., the frequency and amplitude responses, are beyond the scope of this paper and will be shown in subsequent publications.

**B. Combined topographical and electrical contrast**

Many practical applications of the NSMM inevitably involve measuring samples with considerable surface topography. As shown in the numerical analysis, their near-field microwave images will convolve both topographical and electrical information. Proper design of the tip structure and the inclusion of a height control mechanism should, in principle, separate the two contrasts by maximizing one over another. For the current design, the microwave images provides important semi-quantitative information on the material properties since the sample topography can be easily obtained by other methods like SEM, AFM, or simply conventional light microscopy.

Fig. 6 shows a collection of the NSMM images for nano-particles and nano-wires. To demonstrate the resolving power to nanometer-size particles, samples with $Pb(Zr,Ti)O_3$ (Fig. 6(a)) or Au nano-particles (Fig. 6(b)) were prepared and imaged. Particles with size less than 100nm can be well resolved in the images. For imaging of 1D structures, the microwave images of $V_2O_5$ nano-ribbons grown on Au particles and $Ag_2Se$ nano-tubes with contact fingers are shown in Fig. 6(c) and (d), respectively. The data again manifest very high resolving power such that wires with a diameter of 80 ~ 100nm are clearly seen in the images. We emphasize that, unlike conventional AFMs, no laser feedback is needed for the operation of the microwave microscope to probe these nano-scale objects.

NSMM techniques are especially attractive for biological applications because of their sensitivity to water and mineral content [1, 16]. Fig. 7 shows the microwave images of the antennal lobe and compound eyes of the fruit fly Drosophila [19]. Transmission electron microscope (TEM) and light microscope images of similar samples are shown in the insets for comparison. Our microscope is capable to resolve the same features, e.g., the cell and synaptic structures [19] in (a) and (b), and the ommatidial clusters [14, 19] in

(c) and (d), as other imaging instruments. We are currently trying to apply the technique to quantitative biological measurements.

## V. Conclusion

We have demonstrated the design and experimental results of a working near-field scanning microwave microscope. Microwave CPW transmission lines were patterned onto a $Si_3N_4$ cantilever interchangeable with AFM tips. A unique feature of our design is that the sensing and excitation electrodes are separated to suppress the common-mode signal. The RF electronics cancels out the remaining common-mode component and amplifies only the change of the tip signal. Numerical analysis using an electrostatic model shows the high sensitivity and spatial resolution of the microscope. The contrast may originate from variation in either sample topography or its electrical properties. Preliminary microwave images on nano-particles, nano-wires, biological samples, and buried insulator structures were obtained by our NSMM. With refined future tip design and reliable height control for the non-contact mode, we believe that this new microwave imager will greatly impact several research fields and find applications in both fundamental physics and applied science.

## Acknowledgement

The research is supported by the seed grant in Center of Probing Nanoscale (CPN), Stanford University, with partial support from Agilent Technologies, Inc. The cantilevers were fabricated by A.M. Fitzgerald and B. Chui in A.M. Fitzgerald & Associates, LLC, San Carlos, CA. The Pb(Zr,Ti)O$_3$ nano-particle sample was provided by Dr. P. McIntyre's group in the Department of Materials Science and Engineering, Stanford University. The samples with Au nano-particle, V$_2$O$_5$ nano-ribbons, and Ag$_2$Se nano-tubes were provided by Dr. Y. Cui's group in the Department of Materials Science and Engineering, Stanford University. The biological samples were provided by Eric Hoopfer and Chris Winter in Dr. L. Luo's group in the Department of Biological Sciences, Stanford University. The authors acknowledge technical support from Agilent Technologies.

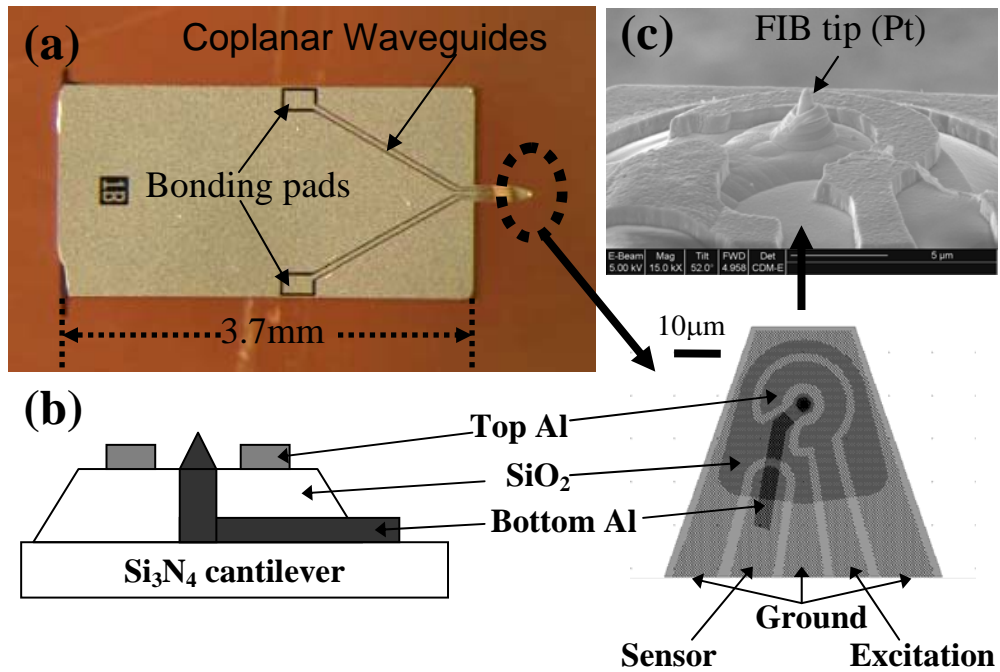

Figure 1. (a) Micro-fabricated cantilever with coplanar waveguides patterned on the top. (b) Schematics of the top view (left) and the side view (right) of the cantilever tip. All electrodes are labeled in the figure. (c) Scanning electron microscope (SEM) image of the Pt-tip formed by focused-ion beam (FIB) deposition.

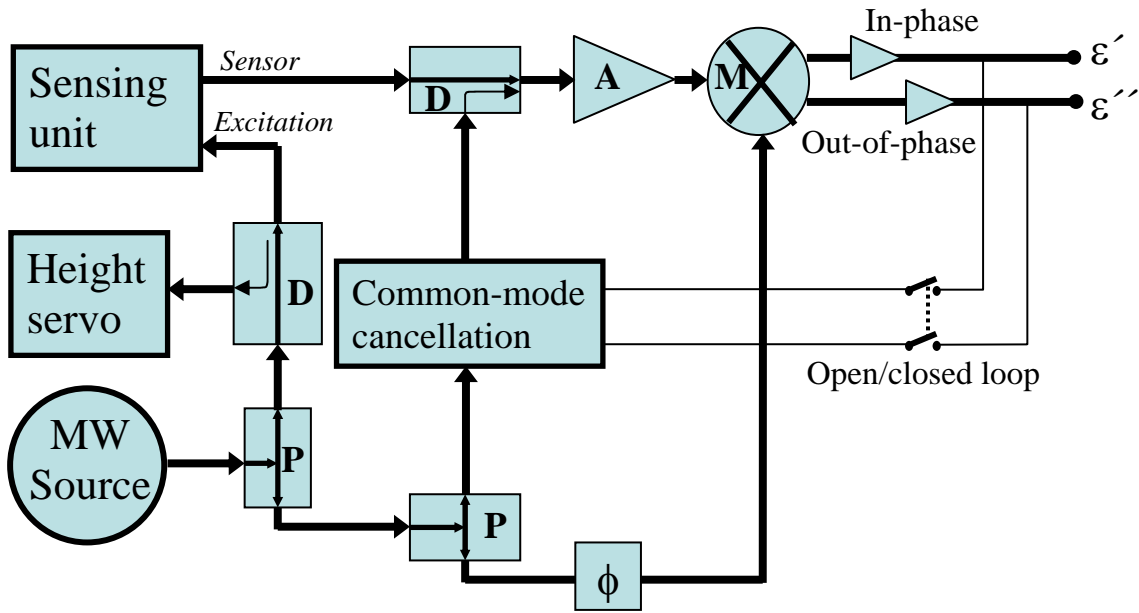

Figure 2. Block diagram of the microwave circuitry. The microwave components are abbreviated as follows, P — Power splitter; D — Directional coupler; A — Amplifier; M — Mixer; ϕ — Phase shifter. Small triangles represent DC amplifiers.

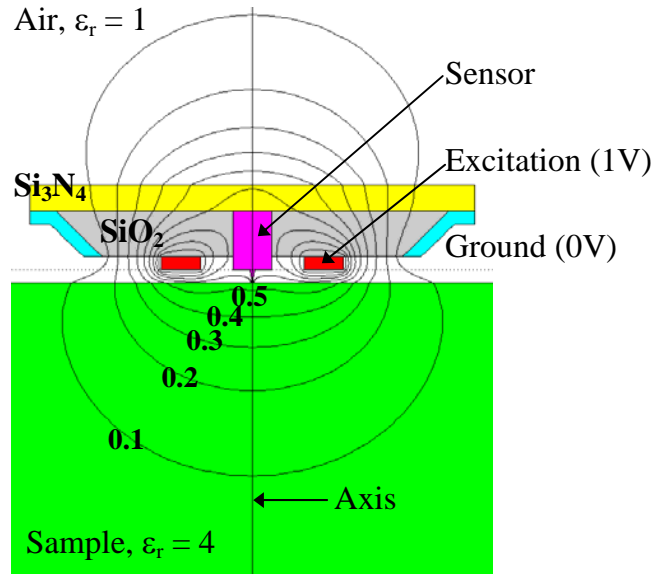

Figure 3. FEMLAB results of the equipotential contour when a sample with $\varepsilon_r = 4$ is in contact with the Pt tip. The 2D axisymmetric electrostatic model was used for the simulation and the excitation voltage is set to 1V.

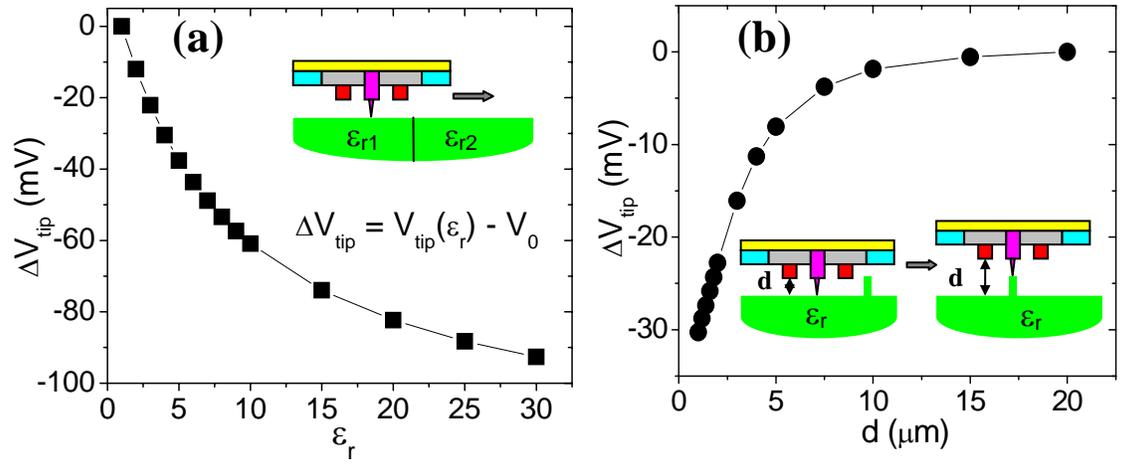

Figure 4. (a) FEMLAB results of the electrical response $\Delta V_{tip} = V_{tip}(\varepsilon_r) - V_0$ as a function of the sample dielectric constant. This response provides the electrical contrast for a flat surface, as shown in the inset. (b) Simulation results of the tip signal when approaching a $\varepsilon_r = 4$ sample. Here the spacing is measured from the excitation electrode. The inset shows how topographical contrast occurs in the contact mode.

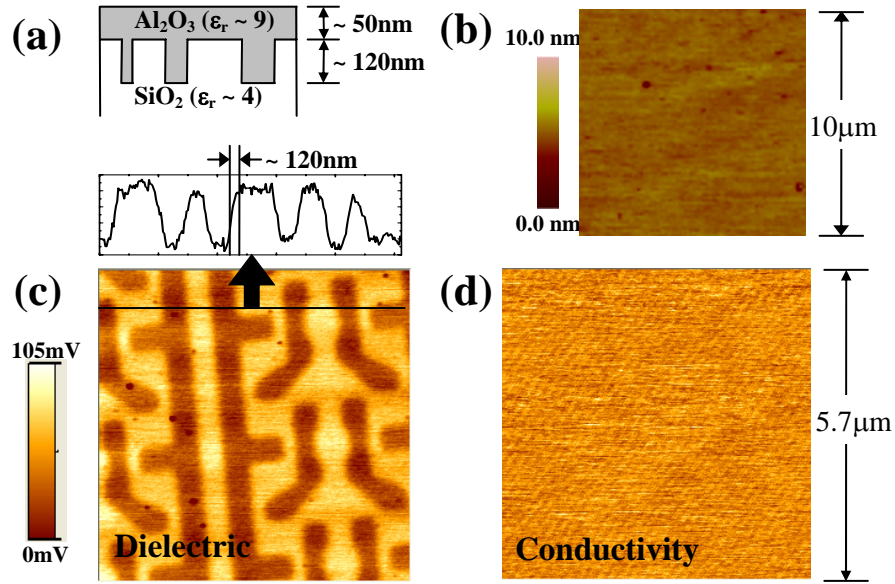

Figure 5. (a) Sample structure -- a layer of $Al_2O_3$ is sputtered onto a $SiO_2$ sample and then the surface is polished. (b) AFM image of the flat surface. (c) and (d) NSMM images from the two output channels in an area of 5.7μm × 5.7μm. Contrast only appears in the dielectric channel and vanishes in the conductivity channel. A line cut of the dielectric channel is also shown and the spatial resolution of the microscope is about 120nm.

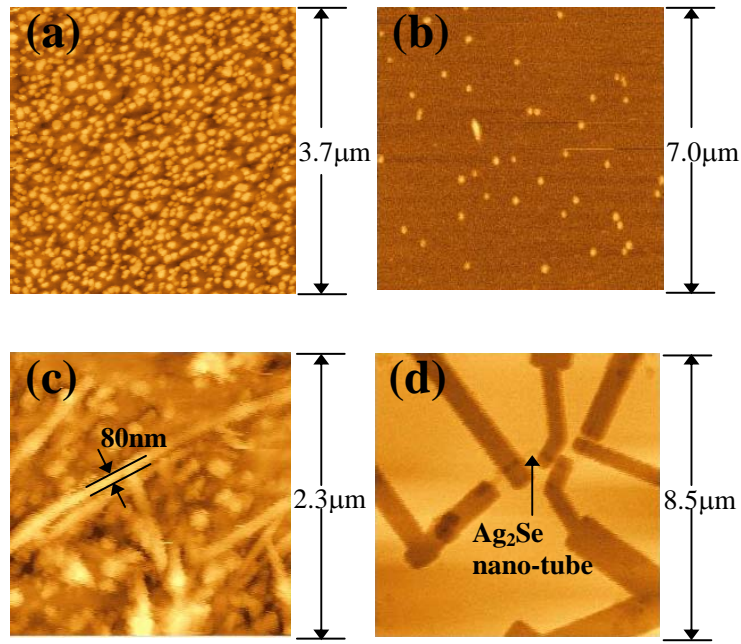

Figure 6. Microwave images convolving both topographical and electrical information. (a) Pb(Zr,Ti)O$_3$ nano-particles. (b) Au nano-particles, (c) V$_2$O$_5$ nano-ribbons. (d) Ag$_2$Se nano-tubes.

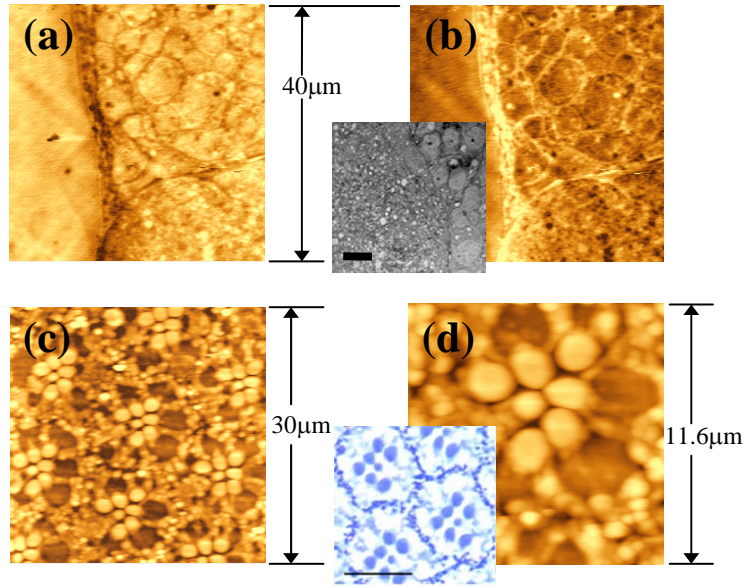

Figure 7. Microwave images of biological samples. (a) In-phase and (b) out-of-phase images of the cell and synaptic structures of an antennal lobe of the Drosophila brain. The inset shows the TEM image of a similar sample. (c) The Drosophila compound eyes and (d) a zoom-in (11.6μm × 11.6μm) view of the same sample, showing clearly the ommatidial clusters. The inset shows the light microscope image. Both scale bars in the insets are 10μm.